\documentclass[%
 reprint,
 amsmath,amssymb,
 aps,
showkeys
]{revtex4-2}

\usepackage{graphicx}
\usepackage{dcolumn}
\usepackage{bm}
\usepackage{multirow}

\begin{document}

\preprint{APS/123-QED}

\title{Quantum Circuit Engineering for Correcting Coherent Noise}

\author{Muhammad Ahsan}
\email{ahsan@alumni.duke.edu}
\affiliation{
	Department of Mechatronics and Control Engineering,\\
	University of Engineering and Technology Lahore
}

\author{Syed Abbas Zilqurnain Naqvi}%
\affiliation{
	Department of Mechatronics and Control Engineering,\\
	University of Engineering and Technology Lahore
}

\author{Haider Anwar}%
\affiliation{
	Department of Computer Science, Namal Institute 
}



\date{\today}

\begin{abstract}
Crosstalk and several sources of operational interference are invisible when qubit or a gate is calibrated or benchmarked in isolation. These are unlocked during the execution of full quantum circuit applying entangling gates to several qubits simultaneously. Unwanted Z-Z coupling on superconducting cross-resonance CNOT gates, is a commonly occurring unitary crosstalk noise that severely limits the state fidelity. This work presents (1) method of tracing unitary errors, which exploits their sensitivity to the arrangement of CNOT gates in the circuit and (2) correction scheme that modifies original circuit by inserting carefully chosen compensating gates (single- or two-qubit) to possibly undo unitary errors. On two vastly different types of IBMQ processors offering quantum volume 8 and 32, our experimental results show up to 25\% reduction in the infidelity of $[[7, 1, 3]]$ code $|+\rangle$ state. Our experiments aggressively deploy forced commutation of CNOT gates to obtain low noise state-preparation circuits. Encoded state initialized with fewer unitary errors marks an important step towards successful demonstration of fault-tolerant quantum computers. 
\end{abstract}

\maketitle


\section{\label{Introduction} Introduction}
	

In state-of-art quantum processors, two-qubit gates are at least order of magnitude noisier than their single-qubit counterparts and limit the fidelity of quantum circuit~\cite{Superconducting_SOP,Iontrap_CNOT_noisier,Silicon_CNOT,Silicon_Single}. Higher operational inaccuracy is not the only bottleneck of state fidelity, action of CNOT gates sequence adds to several context-dependent noise sources including crosstalk~\cite{crosstalk_characterize}, coherent/systematic errors~\cite{IBM_Coherent_detection, coherence_qc}, correlated errors~\cite{correlated_noise_mit} and non-markovian bath~\cite{Non-Markovian_IBM,Ahsan_QIC}. These are some examples of unforeseen errors ~\cite{Circuit_errors} mostly unfolding during the execution of quantum circuit. 
Such circuit-level errors are less visible in the individual gate calibration usually performed prior to the circuit run.  
Several recent studies illustrate prevention~\cite{software_mit_crosstalk,noise_aware,CNOT_commutation,MUQUT,just_in_time,Rod_ibm}, hardware mitigation~\cite{zhao_crosstalk_mit,correlated_noise_mit,DD_Crosstalk,IBM_Rot_echo,CCNOT_AIP} and software mitigation~\cite{ZNE1,ZNE2,ahsan2020} of circuit-level noise. Unfortunately, in the presence of large number of uncorrected errors, it remains unclear how can higher fidelity CNOT gates yield proportionally higher fidelity quantum circuit. 

\begin{figure}
	\includegraphics[width=\columnwidth]{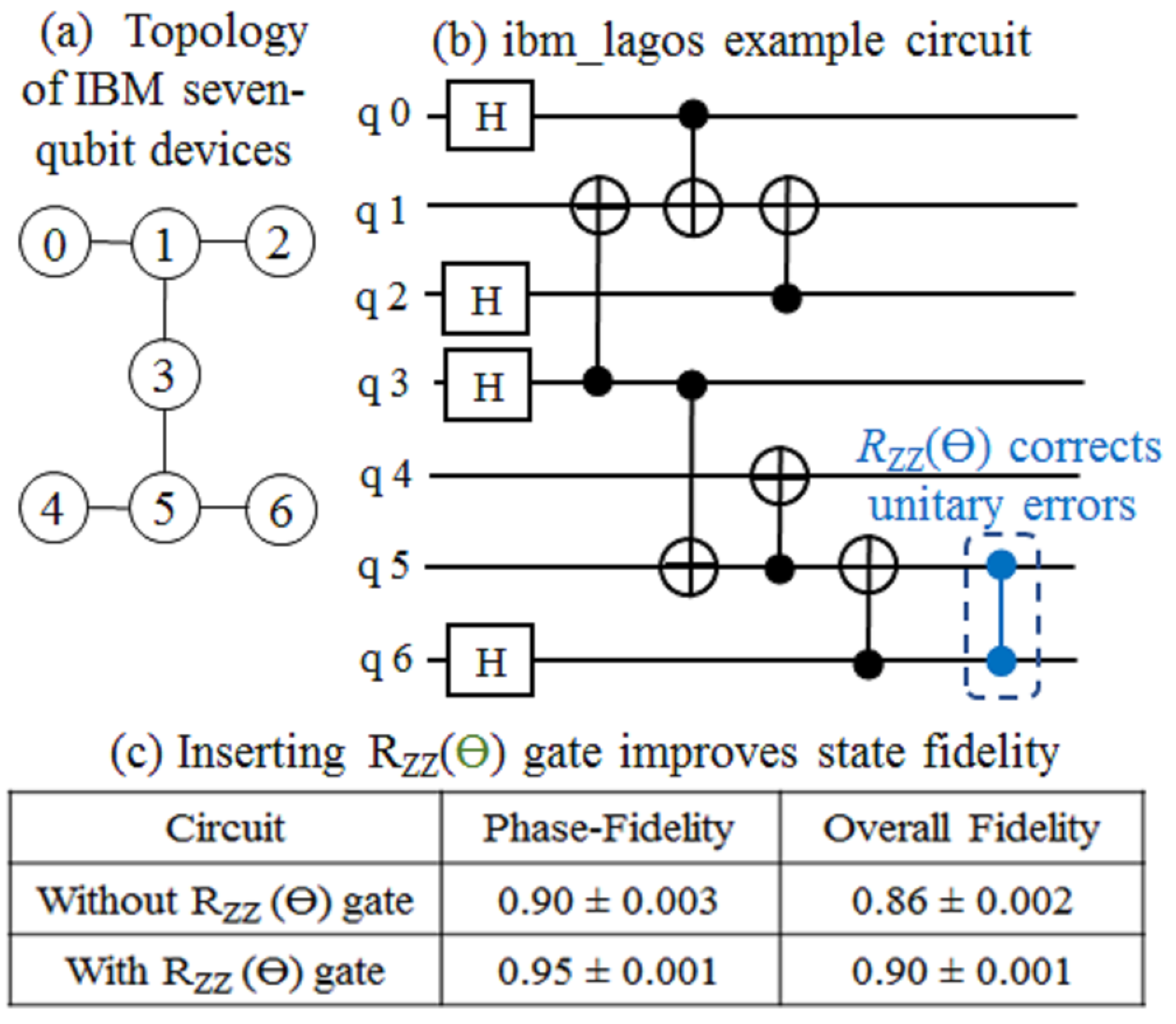}
	\caption{\label{Intro} An example of quantum circuit engineering. Noise compensated (b) state-preparation circuit, containing $R_{ZZ}(\theta)$ gate prepares seven-qubit fully entangled state with higher fidelity than the uncompensated circuit (without $R_{ZZ}(\theta))$. The circuit was run on ibm\_lagos, whose topology, same as that of other seven-qubit devices such as Casablanca and Jakrta, is shown in (a). Table in (c) compares fidelities with and without $R_{ZZ}(\theta)$. Note that $\theta = -\pi/3.5$ in this experiment.}
\end{figure}

\begin{figure*}
	\includegraphics[width=\textwidth]{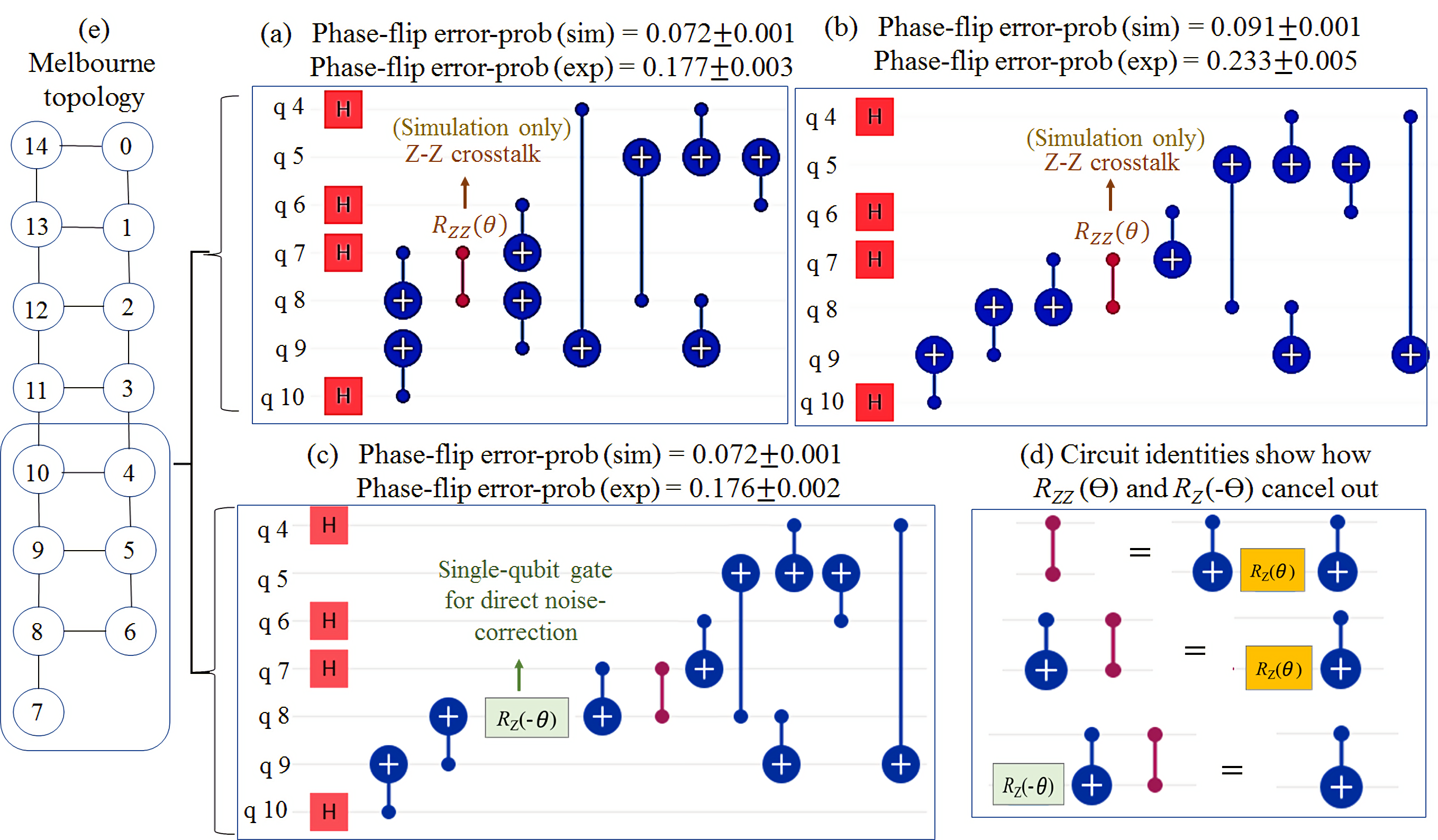}
	\caption{\label{POC} The contextual influence of Z-Z crosstalk on phase-flip error probability $p_z$ and how it can be corrected. Circuits (a) and (b) prepare $[[7, 1, 3]]$ $|+\rangle$ state with slightly different sequence of CNOT gates, with crosstalk on CNOT (q6, q8) in both cases. However, simulation (sim) and experiment (exp) results show that (b) has 20\% higher $p_z$ than (a). Note that crosstalk $R_{zz}(\theta = -\pi/3.5)$ was added to the circuit only in simulation. In circuit (c), noise correcting single-qubit Z-rotation gate $R_Z(\theta = -\pi/3.5)$ cancels crosstalk in (b) by utilizing the circuit identities in (d). The topology of quantum hardware, the ibmq\_melbourne (Melbourne) device is displayed in (e).}
\end{figure*}

\begin{figure*}
	\includegraphics[width=\textwidth]{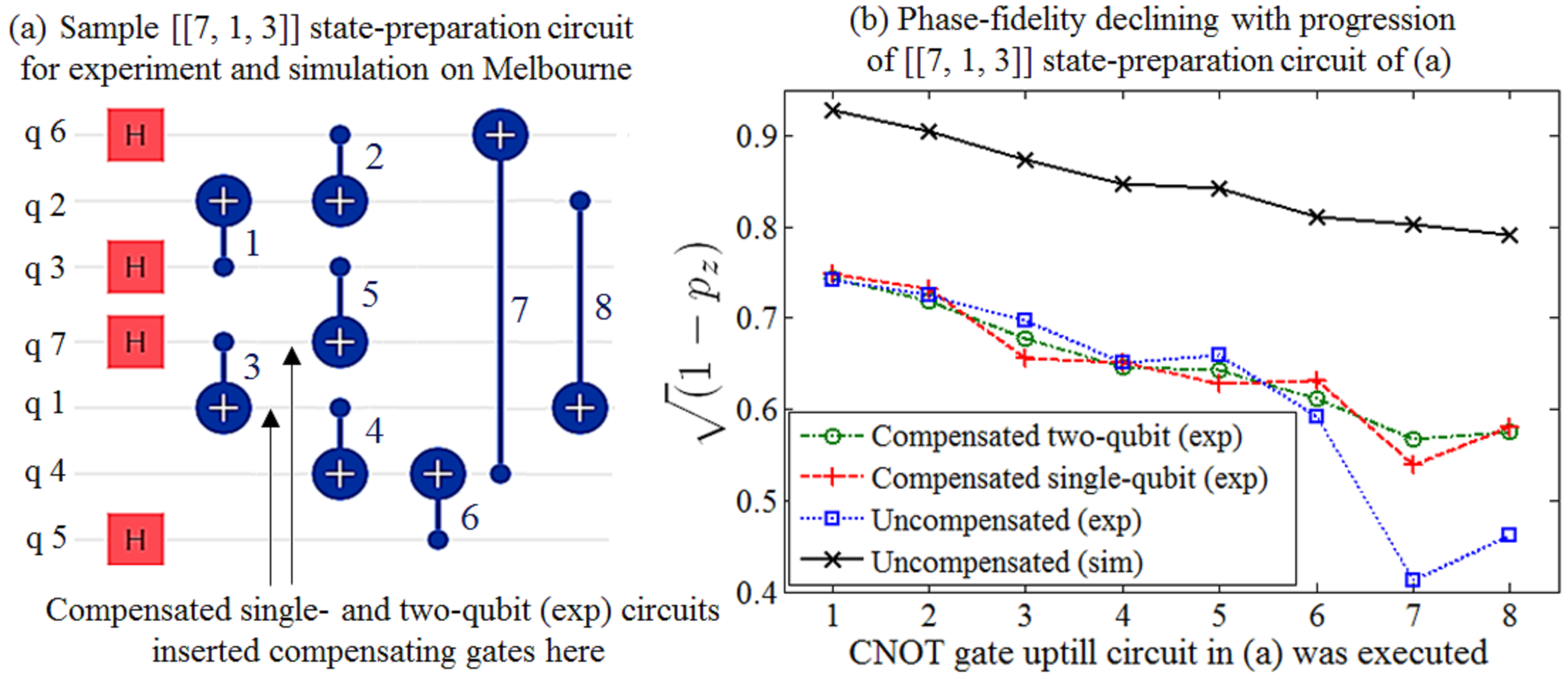}
	\caption{\label{Melbourne} Tracing unitary errors in $[[7, 1, 3]]$ code $|+\rangle$ state-preparation circuit (a) run on Melbourne. The phase-fidelity plummets at CNOT gate-7, indicating added noise due unitary errors. The compensated circuit elevates the curve, and lowers infidelity by 33\%. Circuit (a) also shows indices of CNOT gates for tracing errors in phase-fidelity curve in (b)}
\end{figure*}

This study illustrates quantum circuit engineering for correcting unwanted Z-Z coupling crosstalk and other unitary errors~\cite{IBM_Rot_echo} pervasive in the superconductor cross-resonance CNOT gates~\cite{cross_resonance1,cross_resonance2}. Consequently, noise-compensated circuits are shown to initialize higher fidelity Steane code~\cite{Steane} graph-state on state-of-art IBM quantum processors. 
One of the main findings of this study reveals the sensitivity of sate fidelity to the placement of CNOT gate with Z-Z crosstalk and it serves one part of our noise probe. Commuting CNOT gates, when infected with crosstalk, may yield non-commuting quantum operations in physical circuit, leaving state-fidelity dependent on execution order of the gates. Figure-\ref{POC} provides proof-of-concept state-preparation circuit mapped to IBMQ device Melbourne known to be highly prone to crosstalk noise~\cite{exp_detect_crosstalk} because of of its denser topology. Our simulation and experimental error analysis shows approximately 20\% change in the phase-flip error probability $p_z$, when two commuting CNOT gates are reordered as illustrated in Figure-\ref{POC}. Here $p_z$ is defined as probability that there is at least one phase-flip error on the qubits when their state is readout by X-basis Steane Measurements. The other part of noise-probe comes from tracing $p_z$ along time axis describing progressing execution of the circuit. Addition of crosstalk may significantly raise likelihood of phase-flip errors and introduce marked deviation from otherwise smoothly decaying decoherence curve. Figure-\ref{Melbourne} highlights a precipice at gate-6 in the experimental phase-flip infidelity $\sqrt(1-p_z)$, which sharply contrasts with simulated decoherence curve without crosstalk or unitary phase-flip errors. Section-\ref{tracing}  contains further elaboration of these results. Once detected, it is possible to cancel Z-Z coupling by inserting compensating gates.

Correction procedure can be direct or indirect, the former is illustrated in Figure-\ref{Intro} and Figure-\ref{POC}. In Figure-\ref{POC} (d) circuit identities, Z-Z coupling on CNOT gate simplifies to a single-qubit rotation about Z-axis. It can be corrected by applying conjugate single-qubit gate at the appropriate circuit location. Indirect cancellation is more subtle and relies on experimental intuition that Z-Z coupling on two CNOT gates in the circuit, can cancel each other. Inserting a compensating two-qubit stabilizer gate of the form: H a; CNOT (a,b); H a; X b into carefully chosen circuit location may introduce opposite angle Z-Z coupling to cancel original crosstalk, in a manner very similar to the direct method. Figure-\ref{Melbourne} evidences this effect in the form of nearly identical compensated circuit infidelity curves; both schemes insert compensating gates at same circuit location and prevent fidelity curve from plummeting at gate-7. Our experiments also highlight significant overall improvement in the Steane state fidelity on both the noisier (quantum volume (QV) = 8) as well as on less noisy (QV = 32) IBM quantum processors. 

There is another interesting dimension of unitary error correction; it overcomes an important performance limiting factors of experimental quantum error-correction. Even if initially uncorrelated, a single-qubit rotation, for example, translates into (1) unwanted two-qubit gate on two encoding qubits or (2) between encoding qubit and ancilla, as they propagate through entangling gates,  inducing imperfect parity-check operation even if their CNOT gates are assumed ideal. In principle, accurate tracing of these errors remains a difficult problem in fault-path counting and threshold estimates~\cite{threshold2}. A state-preparation scheme low in unitary can adequately address this problem and facilitates effective implementation of error-correction in the near-term quantum processors. Further discussion is organized into four sections.  Experiment design details and graph-state circuits can be found in in Section-\ref{Experiments}, results and discussion compose Section-\ref{results}. The summary of relevant prior work can be found in Section-\ref{previous work}, while conclusion constitutes Section-\ref{conclusion}. 

Note that in the terminology of this manuscript, the word \emph{noise correction} means correcting unitary errors by inserting compensating gate(s) to the quantum circuit. Such circuit will be called \emph{noise-compensated} or simply a \emph{compensated} circuit.

\section{\label{Experiments} Experimental Tools and Setup}

Pre-requisite of circuit engineering for noise correction involves designing state-preparation circuits with lower decoherence, thereby allowing unitary errors to influence the state fidelity. IBMQ platform enables preparation of Steane $|+\rangle$ state on the 15-qubit Melbourne and on the seven-qubit Casablanca Jakarta and Lagos processors. Their respective topologies, showing qubit-qubit connectivity can be found in Figure-\ref{POC}(e) and Figure-\ref{Intro}(a). Topologies can be modeled as undirected graph describing device level qubit-qubit connectivity. Edges and vertices represent CNOT gates and their operand qubits, respectively. The logical $|+\rangle$ state encoding requires entangling any seven qubits on the quantum processor. Therefore, in case of seven-qubit processors, only single partition of seven qubits is possible. On the other hand, Melbourne can be partitioned into 15 different ways such that each partition clusters seven qubits in a fully connected graph, Figure-\ref{POC}(e) shows one such partition.  Because of fully connected graph, we call these \emph{local} partitions. The significance of local partition lies is reducing the overhead of noisy swaps for CNOT gate on the non-local qubit operands, which enables more faithful analysis of unitary circuit-level errors. 

\subsection{\label{state-prep circuits} $[[7, 1, 3]]$ State preparation (encoding) circuits}

There are multiple ways to map virtual qubits (i.e. qubits in hardware agnostic circuit) to physical qubits (i.e. qubits in the real processor) for any local partition. Different maps may produce different sequence of CNOT gates, hence different circuits in the physical device. Note that for the remaining discussion, the term \emph{circuit} encapsulates qubit map as well as CNOT gate sequence. To date, the smallest Steane $|+\rangle$ state circuit contains 9 CNOT gates. 

\subsubsection{Zero overhead (9-gate) circuit for Melbourne}
The gate count can only increase when device topological constraints are taken into consideration. There is no map that avoids non-local CNOT gate, necessitating additional entangling gates for physically non-adjacent qubits. However, by leveraging existing circuit optimization tools such as:
\begin{itemize}
	\item Dynamic (re)labeling of qubits~\cite{FaultTolSmall_harper}
	\item CNOT gate commutation~\cite{CNOT_commutation,ahsan2020}
	\item  Alternate three-qubit cat-state  circuit shown in Figure-\ref{FCommute}(b)
\end{itemize}
we have designed circuit which incurs no overhead entangling gate, containing only 9 CNOT gates (9-gate) when mapped to the Melbourne hardware as depicted in Figure-\ref{POC}. Both circuits initializes qubits into the Steane $|+\rangle$ state and swap q7 and q8. The final qubit labels are inconsequential, hence omitted from the figure. Almost 60\% of Melbourne experiments in Figure-\ref{Rz} and Figure-\ref{HCNOT} ran 9-gate circuit with valid reordering of CNOT gates.         

\subsubsection{Negative overhead (8-gate) circuit for Melbourne}
It is possible to further decrease CNOT gate count with the help of another optimization called \emph{forced commutation}, never previously explored to best of our knowledge. The 9-gate circuit in Figure-\ref{FCommute}(b) eliminates a (red colored) CNOT (q7, q1) gate by reordering non-commuting CNOT (q7, q4) and CNOT (q4, q1) gates and invoking circuit identity in (\ref{only_equation}) to obtain simpler circuit of Figure-\ref{FCommute}(c).
 
\begin{eqnarray}\label{only_equation}
\textnormal{CNOT} (a, b) \hspace{0.15cm}\textnormal{CNOT} (b, c) = \nonumber \\ \textnormal{CNOT} (a, c) \hspace{0.15cm} \textnormal{CNOT} (b, c) \hspace{0.15cm} \textnormal{CNOT} (a, b) 
\end{eqnarray}

\noindent Consequent decrease in gate count also reduces circuit depth to 4. The 8-gate circuit comprises around 40\% of Melbourne experiments in Figure-\ref{Rz} and Figure-\ref{HCNOT}. It has provably lower phase-flip error-probabilities than the 9-gate version for the given partition. In certain Melbourne partitions, adequate noise-cancellation can be achieved only in 8-gate version. An example 8-gate circuit is shown in Figure-\ref{Melbourne}.

\subsubsection{Circuits for seven-qubit processors}
\begin{figure*}
	\includegraphics[width=\textwidth]{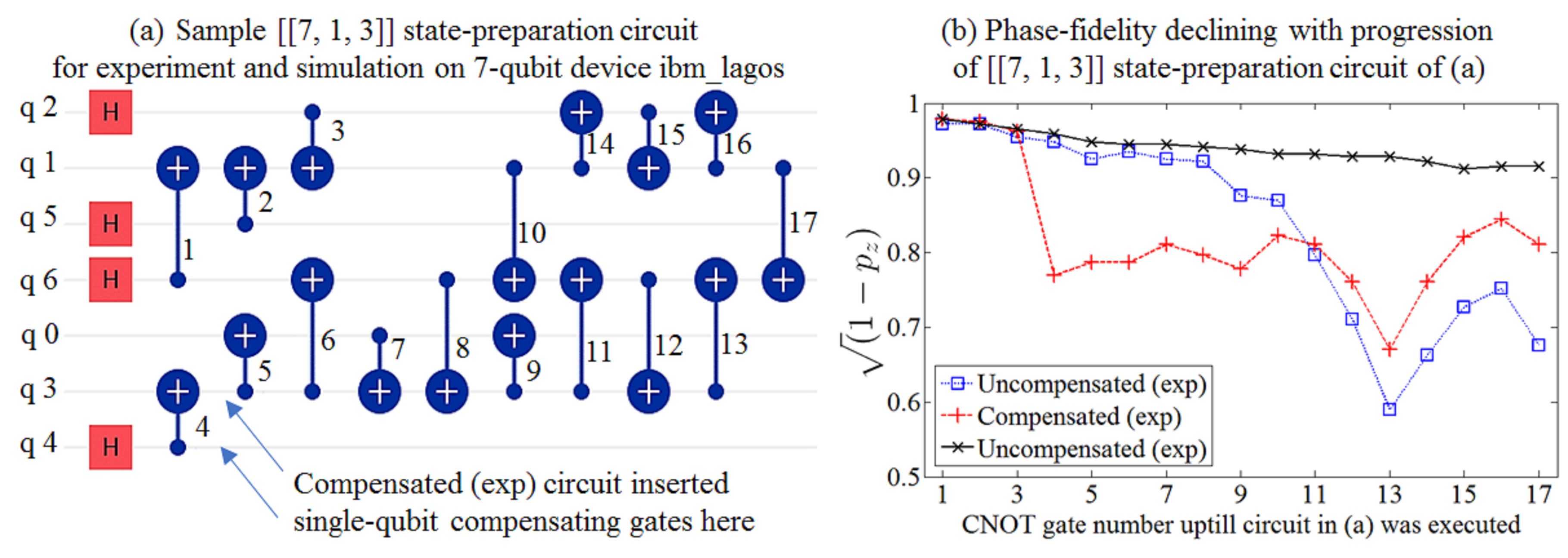}
	\caption{\label{Lagos} Tracing unitary errors in $[[7, 1, 3]]$ code $|+\rangle$ state-preparation circuit (a) run on seve-qubit device Lagos. The phase-fidelity plummets at CNOT gate 13, indicating added noise due unitary errors. The compensated circuit elevates the curve, and lowers infidelity by 50\%.  Circuit (a) also shows indices of CNOT gates for tracing errors in phase-fidelity curve in (b).}
\end{figure*}

In contrast to Melbourne, seven-qubit processors: casablanca, jakarta and lagos exhibit nearly an order of magnitude lower error-rate processors; these offer four-fold higher Quantum Volume (QV) = 32 and their CNOT gates exhibit average failure probability nearly 5x smaller. On the other hand, higher QV comes at the expense of sparser qubit-qubit connectivity; these devices add several swap gates and double the number of CNOT gates in the physical circuit. Using the same set of gate reduction techniques, we obtained 17- and 18-gate state-preparation circuits. Both are identical except upto one superfluous CNOT gate. Unlike Melbourne 8- and 9-gate versions, we did not notice any meaningful difference in error-probabilities. Error analysis of Figure-\ref{Rz} and Figure-\ref{Fidelity}, uses 18-gate version. On the other hand, error tracing example of Figure-\ref{Lagos}(a) contains 17-gate version.

To summarize, all experimental ran circuits derived by valid reordering of commuting CNOT gates of the example circuits given in Figures-\ref{Melbourne} and Figure-\ref{POC} for Melbourne and Figure-\ref{Lagos} and last two rows of Table-\ref{List} for seven-qubit processors. 

\subsection{Computing error-probabilities}
Define, $p_z$, $p_x$ to be the probability of the phase-flip error and the probability of bit-flip error in the Steane $|+\rangle$ state, respectively. Our experiments computed these probabilities from the experiment readout (Steane Measurement) statistics, although readout part is not shown in the circuit diagrams. To demonstrate lower noise level in the compensated circuits, we present two types of experiments (1) Error-probability ($p_z$  or $p_x$) graphs (2) fidelity bar graphs. The first case utilizes Steane Measurements to compute error-probabilities in a single experiment. Seven transversal Z-basis and X-basis measurement are applied at the end of the circuits for $p_x$ and $p_z$ respectively. Their respective values are computed from the readout statistics as $1-$  fraction of readout results $\in \textnormal{C} = [7, 4, 3]$ classical Hamming codeword space, and $1 -$ fraction of readout results in $\textnormal{C}^\perp$ codeword space. Fidelity computation, on the other hand, measures all 128 stabilizers to construct all 128 x 128 density matrix elements. We used IBMQ qiskit code~\cite{ibm_full_fidelity_code} for the fidelity results which closely compare with our fidelity computation $\sqrt(1-p_x)(1-p_z)$ and validates the correctness of our experimental error analysis.

\subsection{\label{tracing} Tracing unitary errors}
Detecting unitary errors requires an adequate tool of tracing error-probability in the circuit. Error tracing identifies appearance of unitary errors by noting the uncharacteristic decline of $\sqrt{1-p_z}$ curve, for example, appearance of deep valleys. Two such examples are given in Figure-\ref{Melbourne} and Figure-\ref{Lagos} for Melbourne and Lagos experiments respectively. To put things in correct perspective, both figures compare experimental and simulation results so that we can quantify how much circuit level noise has been reduced.  For better understanding, we provide relevant details of simulation noise model as follows. Qiskit Ignis tool~\cite{qiskit_ignis} contains several noise models satisfying CPTP constraints, including device specific noise channel derived from latest calibration data. It employs depolarizing channel to model imperfections in the unitary and non-unitary circuit operations, and amplitude and phase-damping channels for the qubit decoherence. The overall noise model then superimposes all these channels to simulate error-probabilities for the whole circuit. However, because it discounts any circuit-level errors, the simulated error probabilities underestimated real noise. As a result,  error-probabilities obtained from device noise model, can only set the lower limit on $p_z$ and $p_x$ obtained from experimental circuits.  The phase-fidelity curve obtained from the simulation of device-specific noise model, provides a credible reference to quantify the circuit-level noise reduction. More details of qiskit Ignis tool can be found in Ref\cite{qiskit_ignis}.

Error tracing skips a set of CNOT gates to initialize qubits in a \emph{partial} $[[7 , 1, 3]]$ state before Steane Measurements. For this purpose, CNOT gates are ordered according to some establish rule of representing scheduling constraints e.g. dependency graph. We run the state-preparation circuit only upto i-th CNOT gate to collect readout statistics. Tracing $p_z$ for each case of $i \in \{1,2,3, \hdots, n\}$, obtains phase-fidelity curve such as those in Figure-\ref{Melbourne} and Figure-\ref{Lagos}. Here $n$ is the total number of CNOT gates in the circuit. For incomplete circuit, the $p_z$ computation first applies missing CNOT gates in post-processing as reversible XOR gates, before computing $p_z$ =  fraction of readout results $\notin \textnormal{C}^\perp$. X-basis Steane Measurements swap the operands of classical XOR operations since H a; H b; CNOT (a,b); H a; H b = CNOT (b, a). 

While simulation phase-fidelity curve declines smoothly throughout, corresponding experimental curve shows similar trend upto a point of steep fall, followed by resurrection. Our experiments show that noise correction proves effective whenever curve shows similar behavior. Deeper valley enable compensated circuits achiever higher reduction in $p_z$.  In both figures, compensatory gates elevate the curve minima, leading to substantial gain in the state fidelity. With reference to simulation phase-flip fidelity curve, noise correction slashes infidelity by 50\%  and 33\% for Lagos and Melbourne respectively. Therefore, it is evident that the most likely cause of sudden decrease in fidelity is a large unitary error on the gate, possibly Z-Z crosstalk, which can be corrected by appropriate conjugate gate. However,  curve resurrection behavior may be explained along two hypotheses. One possibility lies in attributing revival of fidelity to the non-markovian noise. In this model, a quantum circuit can increase the fidelity of evolving state by recovering qubit coherence previously lost in qubit-environment interaction. The recovery is possible if environment coherence lasts till at least next qubit-environment interaction~\cite{detailed_NonMarkovian,exp_non_markovian4,exp_non_markovian5}. Such environment provide basis for the non-markovian noise model. Second explanation views increasing fidelity as partial cancellation of unitary errors on the gates, for example CNOT gate-7 and CNOT gate-8 in case of Figure-\ref{Melbourne} and similar cancellation may occur in case of CNOT gate-13 to gate-16 in Figure-\ref{Lagos}. Latter hypothesis   is simpler and  more consistent with underlying reason of noise-cancellation in our compensated circuits. That said, preliminary evidence of non-markovian or other relevant noise model opens new avenues of future work.

It is possible to sense unitary errors by reordering commuting CNOT gates, which can cause substantial change in $p_z$. Figure-\ref{POC} illustrates this with the help of an example Melbourne circuit simulation as well as experiment. Two circuits (a) an (b) containing same set of CNOT gates are functionally identical but differ in gate sequence. Circuit (b) dispatches CNOT (q4, q9) to the end and interchanges the order of CNOT (q7, q8) and CNOT (q9, q8). An $R_{zz}(\theta)$ gate simulating Z-Z crosstalk on CNOT (q8, q7) produces different $p_z$. In (a), crosstalk acts trivially on the EPR pair, whereas in (b) it introduces non-trivial correlated phase-flips on q7 an q8 and elevates $p_z$ by 20\%. The error-probabilities are higher in the experiments, yet, $p_z$ still increases by at least 20\%, from 0.177 (a) to 0.233 (b). Therefore, altered gate sequence can sense such unitary errors.

\begin{figure*}
	\includegraphics[width=\textwidth]{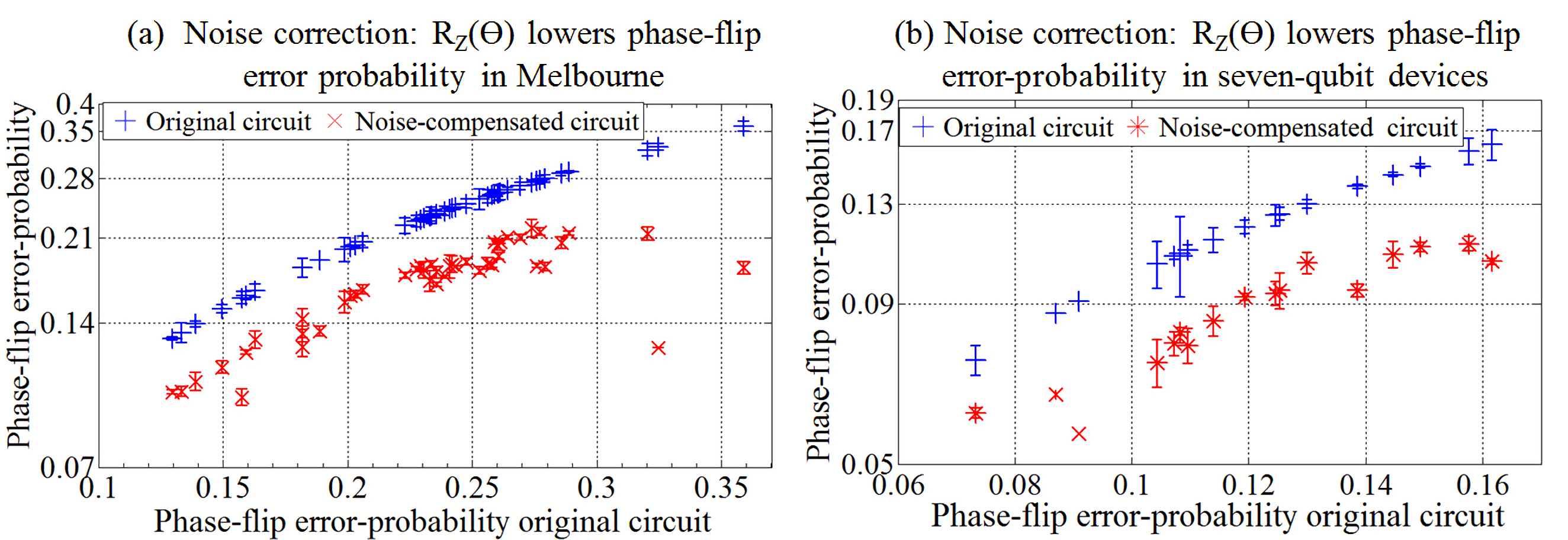}
	\caption{\label{Rz} Compensated circuits lowers phase-flip error probability $p_z$ of $[[7,1,3]]$ code $|+\rangle$ state prepared on Melbourne (a) and seven-qubit devices (b). All compensated circuits insert $R_Z(\theta)$ gate(s) at appropriate locations and lower error-probability by at least 20\%. The value of $\theta$ and circuit location were carefully selected to minimize the noise corrected error-probability. Each data point (original circuit error-prob, compensated circuit error-prob) is vertically aligned on the graph for given abscissa and corresponds to unique circuit (i.e. qubit map and CNOT gate sequence). Melbourne circuits explore noise correction at higher error-rates, whereas, seven-qubit devices circuits shows that it works even at lower error rates. Error bars show 95\% confidence interval.}
\end{figure*}

Based on this example, it is tempting to consider reordering CNOT gate as an obvious path to lower the amplitude of unitary error. While this may work fine for graph-state circuits containing several commuting CNOT gates, it does not constitute a general solution. In other quantum circuits, these gates may not commute in this manner. Therefore, we need a generic noise correction tool whose applicability can be extend to circuits other than graph-state preparation. 

\subsection{Noise Correction}\label{noise correction}

\subsubsection{Melbourne}
An effective noise correction tool should be able to decrease $p_z$ lower than gate reordering. On all local partitions of we found that gate reordering provides upto 20\% change in $p_z$ in the presence of unitary errors. This number was obtained from rigorous analysis of experimental results 9-gate and 8-gate versions of the circuit for Melbourne and 17-gate and 18-gate versions of circuits for seven-qubit processors. Nearly 45,000 circuits were executed, courtesy IBM Quantum Researchers Program, for data collection and analysis. Figure-\ref{POC} shows that while error compensating single-qubit gate $R_z$($\theta$) achieves $20\%$ lower $p_z$ in simulation, the corresponding experiment attains higher reduction---nearly 24\%---in the error-probability. This underpins one of the main contribution of this work. 

Yet $R_z(\theta)$ is not the only route to correct unitary errors, we have found that other single-qubit gates, although not being exact conjugates of unitary errors, can be nevertheless just as effective in certain cases. We omitting their details in the interest of more interesting results. A two-qubit entangling gate of the form H a; CNOT (a, b); H a X b$:=$ HCNOT, inserted into the circuit location wherein it stabilizes the evolving state, can also undo unitary errors in a manner similar to that of $R_z$($\theta$). It somewhat contradicts intutition developed in Figure-\ref{POC} showing Z-Z cross-talk cancellation necessitates non-stabilizer (conjugate) rotations about Z-axis. However, it is not difficult to explain how HCNOT gate replicates cross-talk noise-cancellation of $R_z$($\theta$). Earlier we described that Z-Z couplings on two different CNOT gates can interfere destructively. The HCNOT gate, like other entangling gates, is also noisy, however, it is possible to manipulate its noise to cancel errors and improve state fidelity. A non-ideal HCNOT gate can introduce reverse rotation (e.g. $R_z$($-\theta$) or coupling ($R_{zz}$ ($-\theta$)) to become noise correcting gate. The location of HCNOT gate becomes crucial nonetheless; it must be inserted at a circuit location to ensure that it acts trivially on the ideal state. Our experiments show that HCNOT based error-correction is more effective in Melbourne whose gates have are at least an order of magnitude higher error-rates. This is not a surprising result, after all, we wish to counter noise with noise!

\begin{figure*}
	\includegraphics[width=\textwidth]{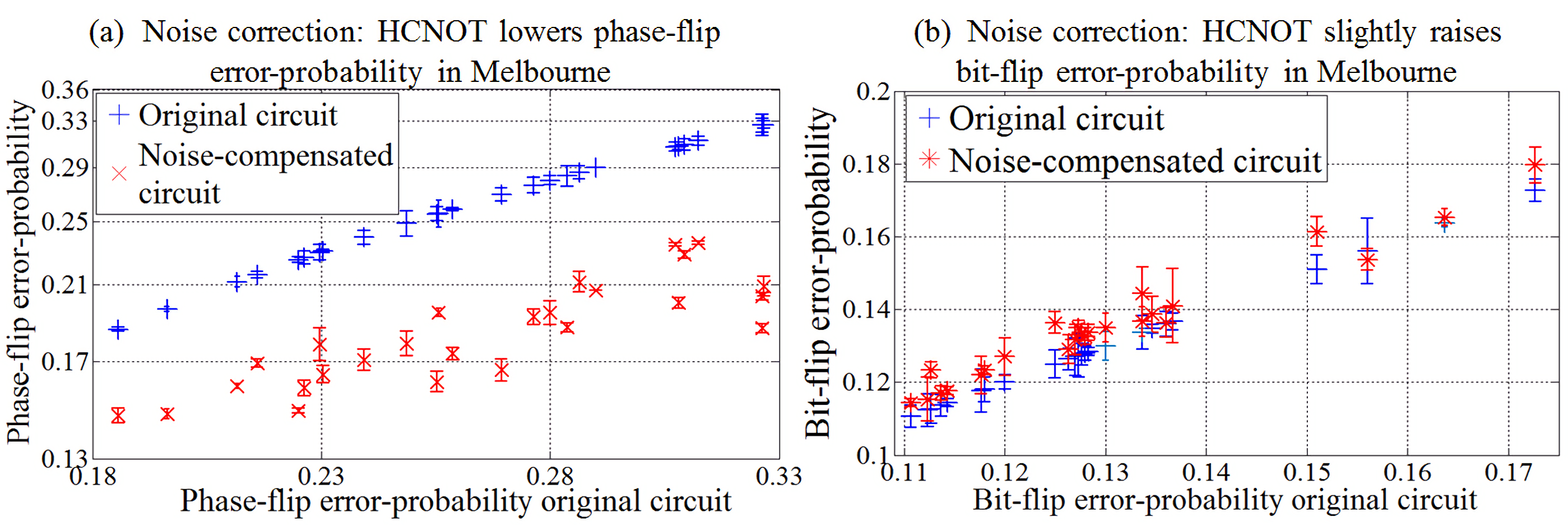}
	\caption{\label{HCNOT} Melbourne compensated circuits lowers phase-flip error probability in (a) while nominally increasing corresponding bit-flip error-probability (b) of $[[7,1,3]]$ code $|+\rangle$ state. All compensated circuits insert HCNOT gate(s) at appropriate locations and lower error-probability by at least 20\%. The circuit location for the gate insertion was carefully selected to minimize the noise corrected error-probability. Each data point (original circuit error-prob, noise-compensated circuit error-prob) is vertically aligned on the graph for given abscissa and corresponds to unique circuit (i.e. qubit map and CNOT gate sequence). These and circuits of Figure-\ref{Rz}(a), exhibit 20\% or higher decrease in phase-flip error-probability on all local partitions of Melbourne. Error bars show 95\% confidence interval.}
\end{figure*}

We further illustrate noise correction with HCNOT for an example circuit in Figure-\ref{Melbourne}. Phase-fidelity curve nosedives at gate-7 indicates unitary error on CNOT (q4, q6). 
With the help of qiskit qasm simulator, we systematically shortlisted circuit locations wherein single or multiple insertions of HCNOT stabilize the evolving state, hoping that in real experiment, some form of noise correcting unitary would accompany HCNOT gate. 
Among feasible candidate locations, our simulations revealed that a combination of HCNOT (q7, q1) and $R_{zz}$ ($\theta = -\pi/3.5$) on qubits q4 and q1, inserted at circuit location shown by arrows, best cancels error on CNOT (q6, q5). At the same location, HCNOT gate with tensor product of $R_z$($\theta = -\pi/7$) rotations on q4 and q1, also work equally well. 
Interestingly, in real Melbourne experiment, inserting HCNOT works as expected, raising phase-fidelity curve at CNOT (q4, q6), as does $R_z$($\theta$) curve. Close resemblance between the two experimental curves of Figure-\ref{Melbourne}(b) further strengthens the hypothesis that HCNOT adds noise correcting gates to mimic phase-flip curve of $R_z$($\theta$).  

\subsubsection{Seven-qubit processor}
Further experimental evidence of noise cancellation can be found in the phase-fidelity curve of circuit run on recently unveiled seven-qubit Lagos processor. It exhibits lowest error-rates among publicly accessible seven-qubit computing chips. Figure-\ref{Lagos}(b) shows that adding $R_z$($\theta$) compensating gate to the circuit despite initially lowering the curve, eventually rebounds and successfully sustains phase-flip fidelity significantly higher than the uncompensated circuit, in the region of gate-13 and onward. Initial decline is less visible in Melbourne experiments probably due to high decoherence rate dictating phase-fidelity in the early stages of the circuit. Lagos, on the other hand, has much longer qubit coherence times; its circuit suffers lesser decoherence, allowing unitary noise to strongly influence fidelity. The impact of unitary noise is visible only in the latter stages of Melbourne circuits. A quick calculation in Figure-\ref{Lagos} shows that the initial 20\% deficit of fidelity (0.96 $\rightarrow$ 0.76) at gate-4 in the compensated circuit, transforms into 19\% gain in fidelity (0.68 $\rightarrow$ 0.81) by the end of the circuit. This remarkable symmetry provides much cleaner evidence of noise cancellation.

\section{Experimental Results and Discussion}\label{results}

We now discuss phase-flip error-probability experiment results for various Melbourne and seven-qubit processor circuits containing single partition. The goal is to (1) show at least 20\% lower $p_z$ with compensated circuit and (2) obtain trend of noise corrected $p_z$ with an overall increase in the noise. Large number of circuits were experimentally explored to satisfy both requirements. Results of Melbourne and seven-qubit processors are distinguishable because of significant difference in topologies, error-rates, circuit sizes and depths.

\begin{figure*}
	\includegraphics[width=\textwidth]{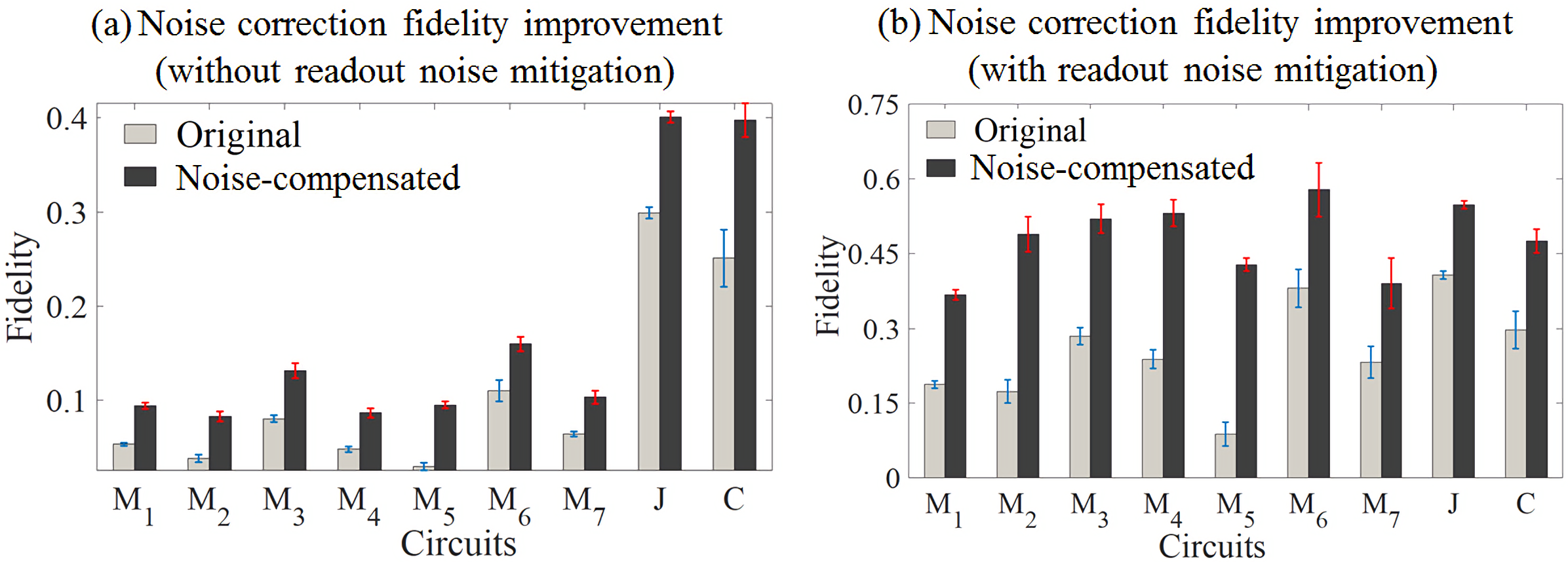}
	\caption{\label{Fidelity} $R_Z(\theta)$ Noise correction increases fidelity of $[[7, 1, 3]]$ code $|+\rangle$ state for selected circuits run on IBM quantum processors; without measurement noise mitigation in (a) and with readout noise noise mitigation~\cite{readout_mit}. Refer to the Table-\ref{List} for details of circuits used in the comparison. Error bars show standard deviation.}
\end{figure*}

\subsection{Reducing phase-flip errors using single-qubit compensatory gate Rz($\theta$)}\label{Rz theta}
Figure-\ref{Rz} compares experimental $p_z$ of uncompensated (original) and corresponding noise compensated circuit containing $R_z$($\theta$) gates by plotting ordered pairs $p_z$ (original), $p_z$ (noise-compensated)) w.r.t $p_z$ (original) for given circuit. The two error probabilities in the pair are vertically juxtaposed, that is, these are meant to be compared along the graph ordinate.  This setting enables comparison over wide range of $p_z$ (original) describing noise-levels available in experimental circuit space. We note that noise correction decreases phase-flip error probability in the noise compensated circuits. In fact, all order pairs lower error-probability by at least 20\%. In some cases, the decrease can be 25\% or even higher. Melbourne experiments (Figure-\ref{Rz}(a)) examines the efficacy of noise correction at comparatively higher noise levels whereas, seven-qubit processors (Figure-\ref{Rz}(b)) highlight its performance at lower noise levels. Compensated circuit are adequately effective in both cases. On the other hand, adding $R_z$($\theta$) does not change bit-flip error probability $p_x$ except for negligibly small statistical fluctuations. When plotted on the graph, data points of the two circuits were indistinguishable, hence excluded from the discussion.

\subsection{Reducing phase-flip errors using two-qubit compensatory gate: HCNOT}\label{HCNOT gate}

The HCNOT compensated circuit, similarly lowers phase-flip error probability although at the cost of slight increase in $p_x$. Still, in most cases, the difference remains less than height of error bars representing 95\% confidence interval. Figure-\ref{HCNOT}(a) displays all ordered pairs in which $p_z$ is lowered by at least 20\% in the compensated circuit. Corresponding $p_x$ are compared in Figure-\ref{HCNOT}(b) showing small increase in $p_x$, conserving net decrease in overall error-probability. The compensated HCNOT noise correction represents an indirect form of noise correction and remains exclusive to Melbourne whose entangling gates have high error-rates. Adding noisy stabilizer two-qubit gate can bring new unitary error which cancels the one on the original circuit. We were unable to replicate indirect error-correction in the seven-qubit processors.

\subsection{Quantifying overall noise reduction} \label{overall fidelity}
\begin{figure*}
	\includegraphics[width=\textwidth]{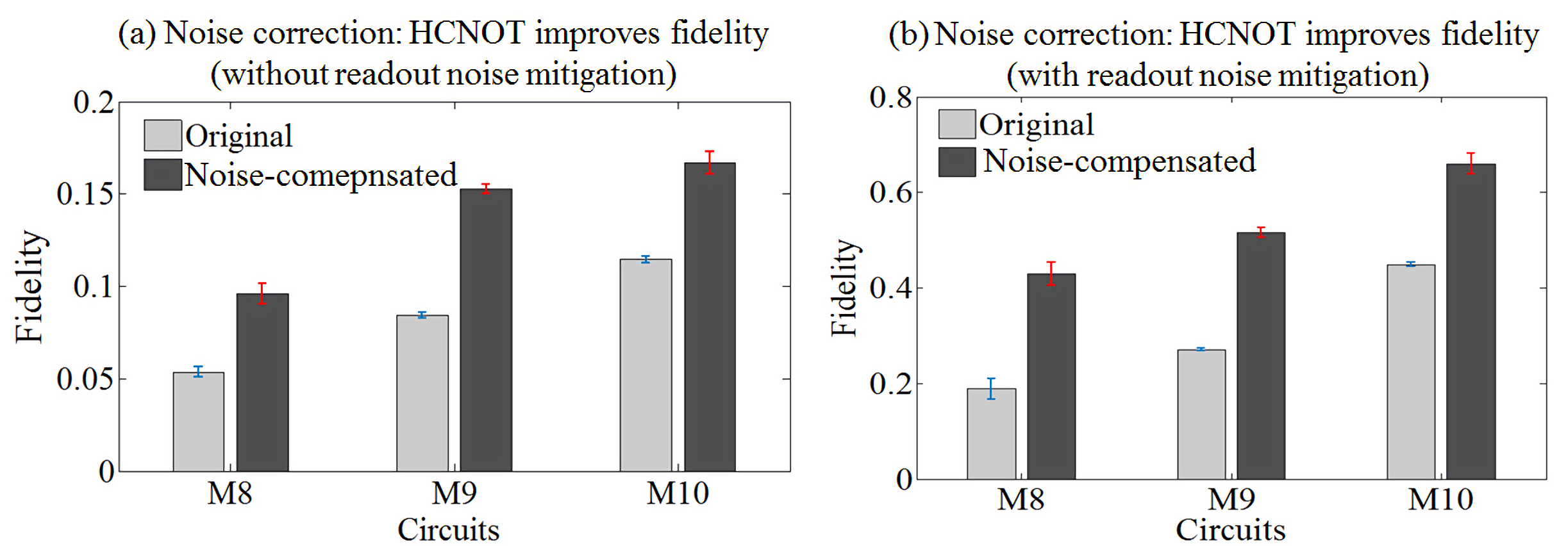}
	\caption{\label{HCNOT_fidelity} HCNOT noise correction increases fidelity of $[[7, 1, 3]]$ code $|+\rangle$ state for selected Melbourne circuits; without measurement noise mitigation in (a) and with readout noise mitigation~\cite{readout_mit} in (b). Refer to Table-\ref{List} for details of circuits used in the comparison. Error bars show standard deviation.}
\end{figure*}

The last set of experiments compute fidelity to show overall noise reduction. One way to experimentally obtain this metric is by using the simplified definition $F = \sqrt{(1-p_z) \times (1-p_x)}$ computing square root of probability that encoding state contains neither bit-flip nor phase-flip during state encoding. The definition also encapsulates the event of no Y-error (bit- and phase-flip) because it decomposes into bit-flip and phase-flip errors upon destructive Steane Measurements. We validated this definition by comparing results with those obtained from more rigorous IBM graph-state fidelity code \cite{ibm_full_fidelity_code}. The comparison showed remarkable similarity between the two set of results. The IBMQ code calculates fidelity from full density matrix description of the graph-state and required 128 experiments per data point. For selected local partitions of Melbourne and qubit map of seven-qubit devices, the fidelity results obtained from IBMQ code are summarized in Figure-\ref{Fidelity} and Figure-\ref{HCNOT_fidelity} for $R_z$($\theta$) and HCNOT compensated circuits, respectively.  Both figures show fidelity improvement with and without measurement noise mitigation~\cite{readout_mit}, for selected circuits of Figures-\ref{Rz} and Figure-\ref{HCNOT}. Qiskit Ignis contains readout noise-mitigation routines using linear filtration: $v = B^{-1}e$. Here B is $2^n \times 2^n$ matrix containing conditional probabilities,  the vectors $v$ and $e$  represent filtered and unfiltered (actual) readout probability distributions. Entries in $B$ matrix are conditional probabilities: \emph{P}(actual\_readout $|$ correct\_readout) and come from separate set of experiments. The random variables \emph{actual\_readout} and \emph{correct\_readout} are seven-bit long string, quantifying the likelihood of obtaining correct result for known (classical) state of qubits. Unitary errors are more pronounced at lower readout noise floor and enable noise correction to better highlight the gain in fidelity.

Bar charts in Figure-\ref{Fidelity} and Figure-\ref{HCNOT_fidelity} juxtapose state fidelities for compensated circuit $R_z$($\theta$)) and original circuits. Define \emph{infidelity} as 1-fidelity to situate these results within the context error-probability graphs. Only the decrease in phase-flip error-probability factors into lower infidelity because bit-flip errors remain uncorrected. In case of HCNOT gate, these may slightly increase infidelity. Still, several compensated circuits achieved 20\% less infidelity for both compensated circuits and on all devices with mitigated readout noise. Overall, these graphs show several compensated circuits lowering infidelities by more than 25\%, and in one circuit $M_5$, the reduction can even reach 35\%, thereby validating the effectiveness of noise correction scheme.

\section{\label{previous work} Previous Work}
Noise cancellation adds to the repertoire of schemes designed to counter errors in Near Term Noisy Intermediate Scale Quantum (NISQ) computers~\cite{NISQ}. At the same time, it features in-situ correction of unitary errors without ancilla overhead---a novel attribute in the context of relevant prior work.  It can also be instrumental to achieve high-fidelity circuit execution on the generation of low decoherence quantum processors. For example, many cloud accessible superconductor quantum computing platforms are converging to heavy hexagonal topology offering quantum volume as high as 64~\cite{IBM_QV_64} by IBM. Improved quantum hardware lowers decoherence which increases the allows unitary errors contribution to state infidelity. Several recent works have addressed various forms of noise and may be broadly characterized as gate-level and circuit-level, though this bifurcation may be less crisp in some cases. They share same underlying theme-- noise suppression, prevention or mitigation.

Gate-level approaches typically rely on advances in pulse shaping~\cite{pulse_shape_leakage}, control~\cite{dynamic_decoupling1,dynamic_decoupling2,dynamic_decoupling3}, dynamically corrected gates~\cite{correlated_noise_mit} and in some cases adding compensating pulses~\cite{IBM_Rot_echo,error_trans2} to decouple principal quantum system state from the environment and cancel unwanted Hamiltonian terms in entangling gate implementation. Recently, dynamic decoupling~\cite{DD_Crosstalk} has been shown to effectively suppress the Z-Z crosstalk noise and improving coherence times. Circuit-level noise approaches can be classified as either pre-processing or post-processing in nature. The former applies hardware calibration data to obtain noise-aware qubit map and gate schedule~\cite{noise_aware,CCNOT_AIP,MUQUT,Rod_ibm}. Gate commutation properties~\cite{ahsan2020,CNOT_commutation} reduce SWAP gate overhead for lower gate count to lower accumulated noise. Just-in-time~\cite{just_in_time} compilation takes experimentally obtained fresh calibration into account for less noisy circuit layout. Detailed noise characterization and intelligent gate scheduling mitigates crosstalk~\cite{software_mit_crosstalk} on 20-qubit IBM quantum processors. 

On the other hand, the post-processing circuit-level schemes modify probability distribution of the circuit readout results such that mean value of an observable of our interest, becomes accurate at the cost of increased variance. This can be achieved either by artificially scaling the error-rates per gate~\cite{ZNE1,ZNE2} with the help slower execution (zero-noise extrapolation) or by carefully depolarizing~\cite{ZNE1} the circuit (probabilistic noise cancellation). These post-processing schemes only improves the estimate of mean value of an observable mapped to readout probability distribution, and do not improve the likelihood of obtaining correct distribution.  Error-correction protocols have been shown to address this shortcoming although at the scale of single logical qubit protected by distance-2 [[4, 1, 2]] code~\cite{FaultTolSmall_harper} as well as distance-3 five-qubit[[5, 1, 3]]~\cite{Five_qubit_FT} and seven-qubit  [[7 ,1, 3]] codes~\cite{seven_qubit_FT}. Very recently, arbitrary error-correction for logical State Preparation And Measurement (SPAM) has been successfully demonstrated for the case of seven-qubit code, achieving SPAM failure-probability of logical qubit lower than its unprotected (physical) counterpart~\cite{seven_qubit_FT}. In any experimental realization of quantum error-correction, high-fidelity encoded state-preparation will be a crucial milestone for the NISQ processors; large number of entangling gates can easily gather enough errors to leave subsequent parity checks operation ineffective~\cite{qec_cycles_fail, mauricio}. Therefore, the correction of unitary errors is crucial to the success of quantum error-correction. 

\section{Conclusion} \label{conclusion}
Tracing and correcting unitary errors pose a challenging and important problem in state-of-art quantum computing platforms. We experimentally demonstrate unitary noise detection and correction on IBM quantum computing devices. We have shown that unitary errors, such as undesirable Z-Z coupling, an be sensitive to the sequence of gates in the physical circuit and cause sudden decrease in fidelity, followed by its gradual recovery, sharply contrasting monotonically declining decoherence curve. The depth of valley in the curve indicates amplitude of unitary noise. Noise tracing requires number of experiments proportional to the circuit size. Correction inserts compensatory gates which can cancel unitary errors either directly or indirectly, both are shown to be equally effective. Detailed experiments highlight performance of noise correction scheme and validated by overall gain the fidelity of [[7, 1, 3]] code $|+\rangle$ state prepared on wide array of IBMQ quantum computing hardware with quantum volume ranging from 8 to 32.

Although our case study structures important details of noise behavior, it also unfolds some interesting questions for the future work thereby. For example, how to efficiently trace unitary noise in the non-stabilizer states. Tracing error-probabilities for circuits containing non-clifford gates will likely need full fidelity computation, hence exponentially large number of experiments. In addition, valleys traced by fidelity curve motivates investigation of non-markovian noise which may leave its signatures in the form of coherence revival. Finally, considering encouraging noise cancellation results, how can we alter noise composition towards a mixture more unitary and fewer non-unitary errors? To a certain extent, increasing qubit coherence times and decreasing operational error-rates have already altered the composition in the novel topologies of quantum processors. However, several unmodeled and unmitigated sources of unitary circuit-level errors expand ample space for quantum circuit engineering.  
\begin{acknowledgments}
The views expressed are those of the authors, and do not reflect the official policy or position of IBM or the IBM Quantum team.  Authors used ibmq\_melbourne, which is one of the IBM Quantum Canary Processors as well as ibmq\_casabalanca, ibmq\_jakarta and ibm\_lagos which are seven-qubit IBM Quantum Falcon Processors. We also acknowledge the access to advanced services provided by the IBM Quantum Researchers Program.

\end{acknowledgments}


\bibliography{ref2}
\appendix
\section{}\label{appendix}

\begin{figure*}
	\includegraphics[width=\textwidth]{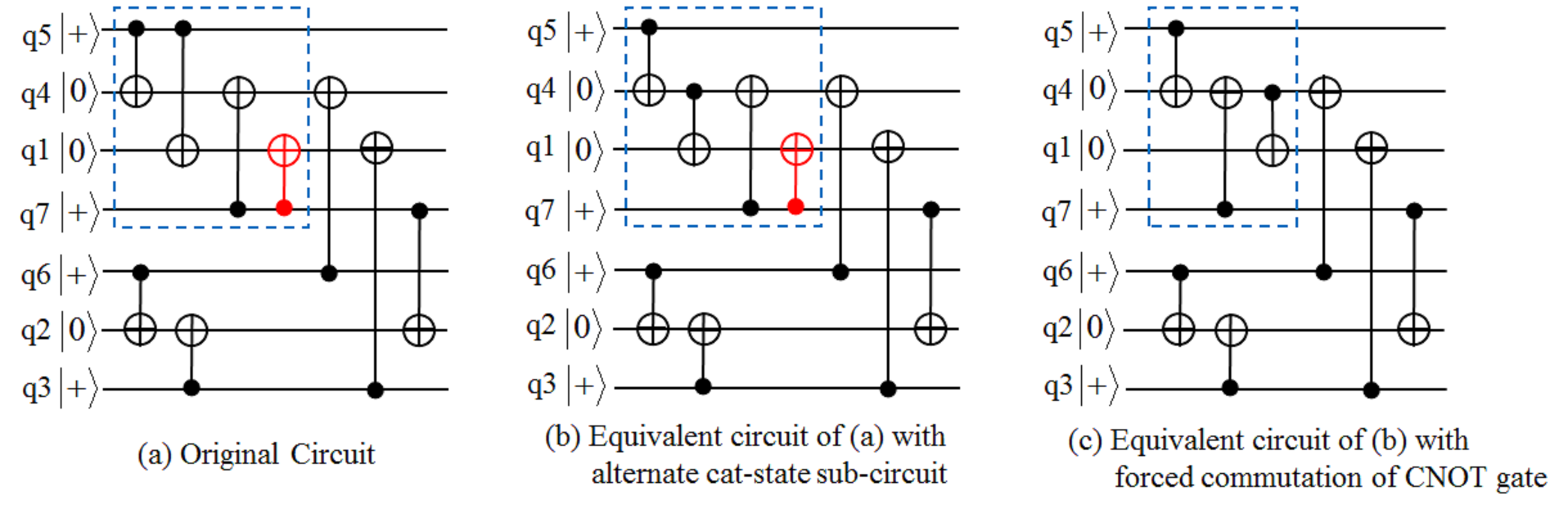}
	\caption{\label{FCommute} Forced commutation technique for decrementing CNOT gate count and circuit depth in the [[7, 1, 3]] code $|+\rangle$ state-preparation circuit. Original circuit (a) can alternatively prepare qubit-5,6 and 7 cat-state by changing the control operand of one of the CNOT gates, as illustrated in (b). Back-commuted CNOT gate on qubits q7 and q4, cancels  CNOT (q7, q1) in (c) by virtue of forced commutation. The technique almost doubled  circuit space for our experiments.}
\end{figure*}

\begin{center}
	\begin{table*}[] \caption{Details of circuits used in Figure-\ref{Fidelity} and Figure-\ref{HCNOT_fidelity}} \label{List}
		\begin{center}
			\includegraphics[width=\textwidth]{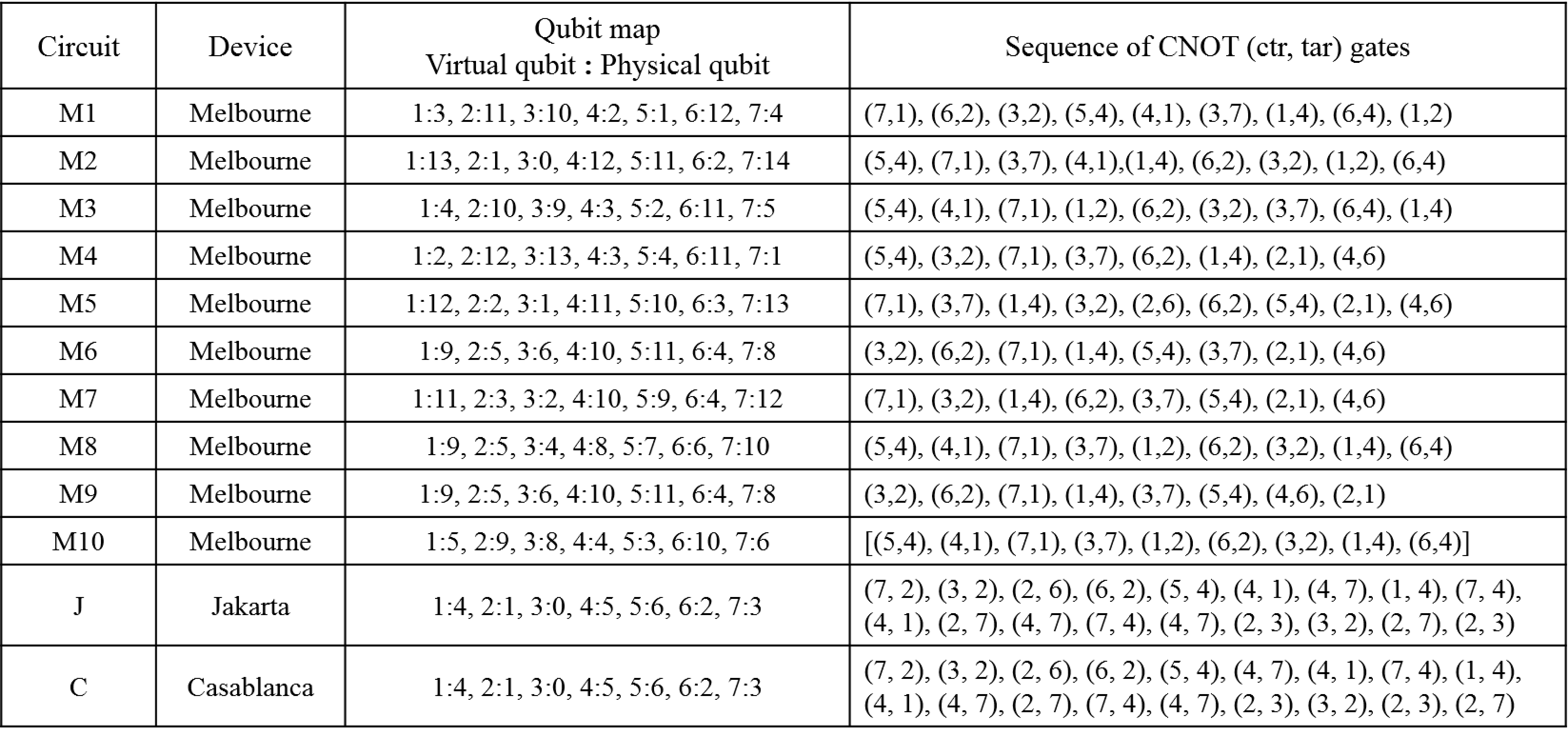}
		\end{center}
	\end{table*}
\end{center}




\end{document}